\documentclass[aps,prl,preprint,showpacs]{revtex4-1}

\usepackage{graphicx}
\usepackage{dcolumn}
\usepackage{bm}
\usepackage[caption=false]{subfig}
\usepackage{float}
\usepackage{subfig}

\begin{document}

\title{Ramsey fringes in a room temperature quantum dot semiconductor optical amplifier}
\author{I. Khanonkin}
\email{ikhanonkin@tutanota.com}
\affiliation{Russell Berrie Nanotechnology Institute, Technion, Haifa 32000, Israel}
\affiliation{Andrew and Erna Viterbi Department of Electrical Engineering, Technion, Haifa 32000, Israel}
\author{A. K. Mishra}
\affiliation{Russell Berrie Nanotechnology Institute, Technion, Haifa 32000, Israel}
\affiliation{Andrew and Erna Viterbi Department of Electrical Engineering, Technion, Haifa 32000, Israel}
\author{O. Karni}
\affiliation{E. L. Ginzton Laboratory, Applied Physics Department., Stanford University, Stanford, CA 94305, USA}
\author{S. Banyoudeh}
\affiliation{Institute of Nanostructure Technologies and Analytics, Technische Physik, CINSaT, University of Kassel, Kassel 34132, Germany}
\author{F. Schnabel}
\affiliation{Institute of Nanostructure Technologies and Analytics, Technische Physik, CINSaT, University of Kassel, Kassel 34132, Germany}
\author{V. Sichkovskyi}
\affiliation{Institute of Nanostructure Technologies and Analytics, Technische Physik, CINSaT, University of Kassel, Kassel 34132, Germany}
\author{V. Michelashvili}
\affiliation{Russell Berrie Nanotechnology Institute, Technion, Haifa 32000, Israel}
\affiliation{Andrew and Erna Viterbi Department of Electrical Engineering, Technion, Haifa 32000, Israel}
\author{J. P. Reithmaier}
\affiliation{Institute of Nanostructure Technologies and Analytics, Technische Physik, CINSaT, University of Kassel, Kassel 34132, Germany}
\author{ G. Eisenstein}
\affiliation{Russell Berrie Nanotechnology Institute, Technion, Haifa 32000, Israel}
\affiliation{Andrew and Erna Viterbi Department of Electrical Engineering, Technion, Haifa 32000, Israel}

\begin{abstract}

The ability to induce, observe and control quantum coherent interactions in room temperature, electrically driven optoelectronic devices is of outmost significance for advancing quantum science and engineering towards practical applications. We demonstrate here a quantum interference phenomena, \textit{Ramsey fringes}, in an inhomogeneously broadened InAs/InP quantum dot (QD) ensemble in the form of a $1.5~mm$ long optical amplifier operating at room temperature. Observation of \textit{Ramsey fringes} in semiconductor QD was previously achieved only at cryogenic temperatures and only in isolated single dot systems. A high-resolution pump probe scheme where both pulses are characterized by cross frequency resolved optical gating (X-FROG) reveals a clear oscillatory behavior both in the amplitude and the instantaneous frequency of the probe pulse with a period that equals one optical cycle at operational wavelength. Using nominal input delays of $600$ to $900~fs$ and scanning the separation around each delay in $1~fs$ steps, we map the evolution of the material de-coherence and extract a coherence time. Moreover we notice a unique phenomenon, which can not be observed in single dot systems, that the temporal position of the output probe pulse also oscillates with the same periodicity but with a quarter cycle delay relative to the intensity variations. This delay is the time domain manifestation of coupling between the real and imaginary parts of the complex susceptibility. 

\end{abstract}


\maketitle

Semiconductor quantum dots (QDs) serve routinely as a viable platform for basic quantum mechanical experiments and detailed understanding of light-matter interactions \cite{natureR,reithmaier2008strong}. The use of QDs for quantum applications such as generation of single photons \cite{michler2000quantum,regelman2001semiconductor,claudon2010highly,benyoucef2013telecom} and entangled photon pairs \cite{benson2000regulated,akopian2006entangled,schwartz2016deterministic,young2006improved}, photon echo based quantum memories \cite{poltavtsev2016photon} and quantum gates \cite{burkard1999coupled} have been demonstrated often and are well documented.

The basic quantum mechanical phenomena, e.g. Rabi oscillation \cite{rabi1939}, Ramsey interference \cite{ramsey1990} etc. enable control over final quantum states. Unlike Rabi oscillation, Ramsey interference involves two time delayed pulses, hence delivers substantially better control over the coherent evolution of the quantum state \cite{ramsey1990}.  Early measurement of Ramsey fringes in a single GaAs QD, which was formed by width fluctuations of a quantum well, was observed using photoluminescence (PL) measurements by Bonadeo \cite{bonadeo1998coherent}. Similar Ramsey-type oscillation in a single InGaAs QD selected from a low density self-assembled system was reported by Toda \cite{toda2000near} and Htoon \cite{htoon2002interplay}. Photocurrent measurements in electric-field-tunable single QD systems were used to demonstrate Ramsey interference by Stufler \cite{stufler2006ramsey} and by Michaelis \cite{de2010coherent}. Two and three pulse photon echo experiments also enabled observation of Ramsey fringes in a single InAs/GaAs QD, as reported by Jayakumar \cite{jayakumar2013deterministic}. A different aspect of Ramsey interference addressed the spin-state of QD which was interrogated using ultrafast optical techniques by Press \cite{press2008complete} by Kim \cite{kim2010ultrafast} and by Lagoudakis \cite{lagoudakis2016ultrafast}. Common to all these reports is the fact that only isolated single quantum dots, held at cryogenic temperatures, were used.

Another class of quantum coherent experiments in semiconductor media were reported by Choi \cite{choi2010ultrafast} who showed Rabi oscillations in a quantum cascade semiconductor laser. Coherent light matter interactions in QD ensembles were studied by Marcinkevicius \cite{marcinkevivcius2008transient} who demonstrated electromagnetic induced transparency in a stack of InGaAs/GaS QDs and by Suzuki \cite{suzuki2016coherent} who used two-dimensional coherent spectroscopy to study an ensemble of InAs/GaAs QDs. Those experiments were also performed at cryogenic temperatures.

We use a different approach to induce and observe quantum coherent phenomena in QD systems in which ultrashort optical pulses excite an electrically driven QD ensembles operating at room temperature. The platform we employ is an electrically driven QD-based semiconductor optical amplifier (QD SOA) which is a long waveguide that provides optical gain. Such experiments have to overcome several hurdles. The first is the short room temperature coherence time which was previously determined from temperature dependent PL linewidth of a single QD by Bayer \cite{bayer2002temperature} and through transient four wave mixing, measured in a GaAs QD SOA by Borri \cite{borri2000time}. Both techniques yield a coherence time from $200$ to $300~fs$. Other difficulties stem from the QD ensemble inhomogeneity and the fact that incoherent interactions such as two photon absorption, and associated Kerr-like effect, take place simultaneously and can mask coherent interactions. Nevertheless, using ~$150~fs$ single excitation pulses, Rabi oscillations were demonstrated by Karni in an InAs/InP QD \cite{karni2013rabi} and by Capua in InAs/InP quantum dash (wire-like nano structures) \cite{capua2014rabi} SOAs employing the cross frequency resolved optical gating (X-FROG) technique and later by Kolarczik in an InAs/GaAs QD SOA \cite{kolarczik2013quantum} using a technique called FROSCH. The role of the gain inhomogeneity and the incoherent interactions was considered by Karni \cite{karni2015nonlinear} as was a demonstration of coherent control to enhance the Rabi oscillations by using shaped excitation pulses \cite{karni2016coherent}. A different type of coherent control with a two-pulse pump-probe-X-FROG was used by Capua \cite{capua2014coherent} to demonstrate cyclical instantaneous frequency variation of the probe pulse in a QD SOA that was biased to the absorption regime. No trace of a periodic intensity oscillation was observed in that experiment probably because of a severe dot inhomogeneity and a significantly reduced interaction of the electromagnetic field with the QDs due to off resonant excitations.

We report here a first observation of Ramsey fringes in an in-homogeneously broadened ensemble of QDs. This is also a first demonstration of Ramsey fringes ever measured in a room temperature semiconductor. A series of experiments using a pump-probe-X-FROG system that includes shaping of the pump pulse demonstrate a clear oscillatory behavior, with a period equal to an optical cycle, of both the intensity and the instantaneous frequency profiles of the probe pulse. The oscillation modulation depth decreases with the nominal input pulse delay thereby enabling direct mapping of the de-coherence and the extraction of the coherence time which was found to be $340~fs$. This value is somewhat larger than those reported in \cite{bayer2002temperature}  and \cite{borri2000time} due to differences between the energy level structures of InP and GaAs based QD systems and the different nature of the measurement techniques. 

An additional major finding described in this paper is an oscillation of the output temporal pulse separation caused by the coupling between the real and imaginary parts of the material susceptibility. The pulse separation oscillates with the same periodicity as the intensity and instantaneous frequency but lags behind them by a quarter of a cycle due to the complex nature of the susceptibility. The variation of the output temporal separation is a direct result of the distributed nature of the pulse propagation along the SOA and cannot be observed in any experiment using a single QD.

The experimental results were confirmed by previously reported comprehensive numerical model \cite{karni2016coherent,capua2013finite}.

\begin{figure*}[htb]
	\centering
	\includegraphics[width=14.6cm]{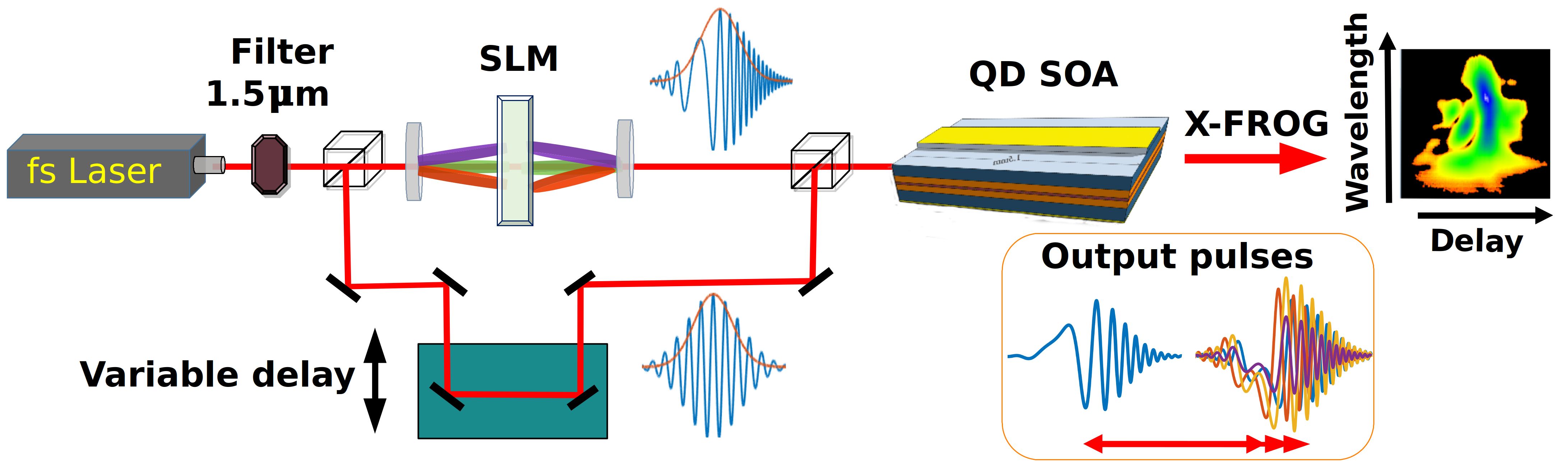}
	\caption{Schematic diagram of the pump-probe-X-FROG system. 150 fs wide pump pulse is
		shaped by an SLM in order to prevent input pulse overlap and to enhance the induction of a
		coherent state to be sensed by the probe. The spatial resolution of the delay line is better than $0.5~fs$.}
	\label{Ramsey_1}
\end{figure*}

The experimental set-up is shown schematically in Fig. \ref{Ramsey_1}. Fiber laser (Toptica FemtoFiber pro) generated pulses with repetition rate of $40~MHz$ are filtered to obtain a $\sim 150 ~fs$ pulses  centered at $1.55~ \mu m$. The pulses are split with one arm traversing a liquid crystal based spatial light modulator (SLM) (Jenoptiks SLM-S640 with a broadband antireflective coating) and the other passing through a highly accurate motorized delay line (PI), the resolution of which is better than $0.5~fs$.

The two pulses are recombined before they are coupled into a QD SOA. The SLM is used to adjust the spectral phase of the pump pulse in order to remove a moderate trailing edge wing of the pulse and prevent any overlap of the input pulses. The input energies of the pump and probe pulses are $35~pJ$ and $20~pJ$, respectively. The pulses are measured at the SOA output using the X-FROG technique \cite{trebino2012frequency}. The amplitude and phase profiles are obtained from the measured data using a phase retrieval algorithm \cite{trebino2012frequency} with a convergence error below $1\%$.

QD based SOA operates at $1.55~\mu m$ wavelength and comprised of six InAs dot layers grown by molecular beam epitaxy in the Stranski-Krastanov growth mode \cite{gilfert2010influence}. Each layer has a density of about  $6 \cdot 10^{10}~cm^{-2}$  with a record uniformity characterized by a photoluminescence linewidth of $17~meV$ at $10~K$ \cite{banyoudeh2015high}. Layer stacking broadens the linewidth due to strain related inhomogeneities. The linewidth of the six-layer structure is nevertheless very narrow, $30~meV$ \cite{banyoudeh20161}, which is also a record. The emission level of electroluminescence spectra increases with bias but the spectral shape remains unchanged \cite{khanonkin2017ultra}. No blue shift due to the plasma effect or a red shift due to self-heating take place. This signifies that the QDs do not saturate and the maximum possible gain is obtainable for each drive current. The large density and the dot uniformity yield an extremely large modal gain of $15~cm^{-1}$ per dot layer which enables superb laser characteristics \cite{banyoudeh2016temperature}. The optical amplifier was realized by coating the end facets of a $1.5~ mm$ long laser with a two-layer dielectric anti-reflection coating yielding modal reflectivities of $0.01 \%$. In the experiments described hereon, the amplifier was biased at $210~mA$ where it exhibits large gain.

\begin{figure}[h]
	\centering
	\includegraphics[width=8.6cm]{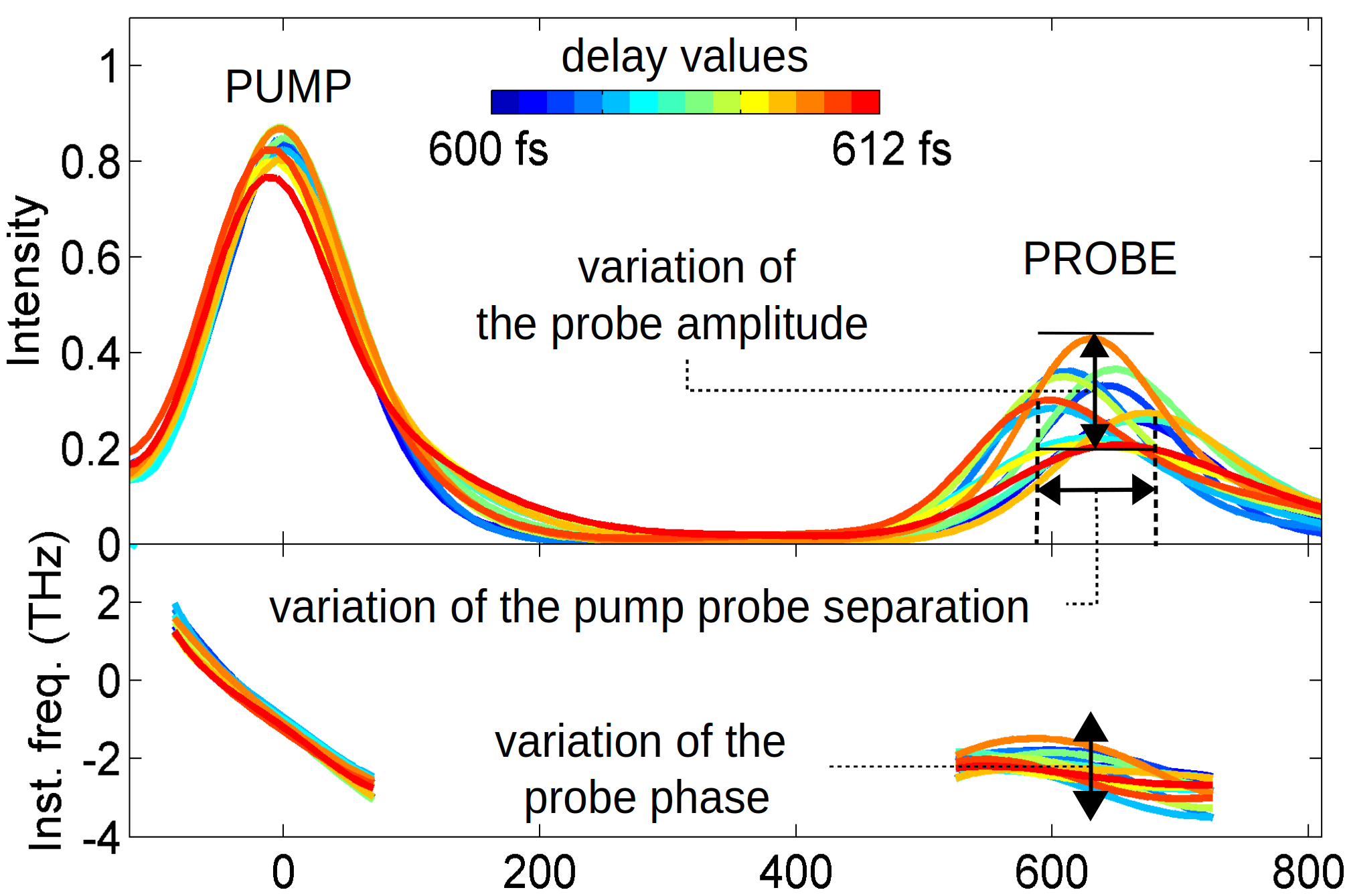}
	\caption{X-FROG traces of the two pulses for a nominal delay of $600~fs$. The upper trace shows intensity profiles for delays of $600$ - $612~fs$. The lower trace shows the corresponding instantaneous frequencies. The pump pulse is delay independent but the probe pulse exhibits clear Ramsey fringes with a periodicity of one optical cycle, $ \sim 5 ~fs$.}
	\label{Ramsey_2}
\end{figure}

The intensity and instantaneous frequency profiles of the output pulses for temporal delays of $600$ to $ 612~fs$ are shown in Fig. \ref{Ramsey_2}. The profiles of the pump pulse are naturally independent of the delay, however, the probe pulse exhibits cyclical intensity variations accompanied by oscillations in the instantaneous frequency profile. The pump pulse sets a quantum state, which is sensed by the probe pulse whose phase, relative to that of the coherent state, varies as the delay changes in $1~fs$ steps (which corresponds to less than a quarter of an optical cycle) resulting in the oscillatory behavior which is a clear indications of Ramsey interference fringes \cite{salour1978quantum}. 

\begin{figure*}[htb]
	\centering
	\includegraphics[width=17.6cm]{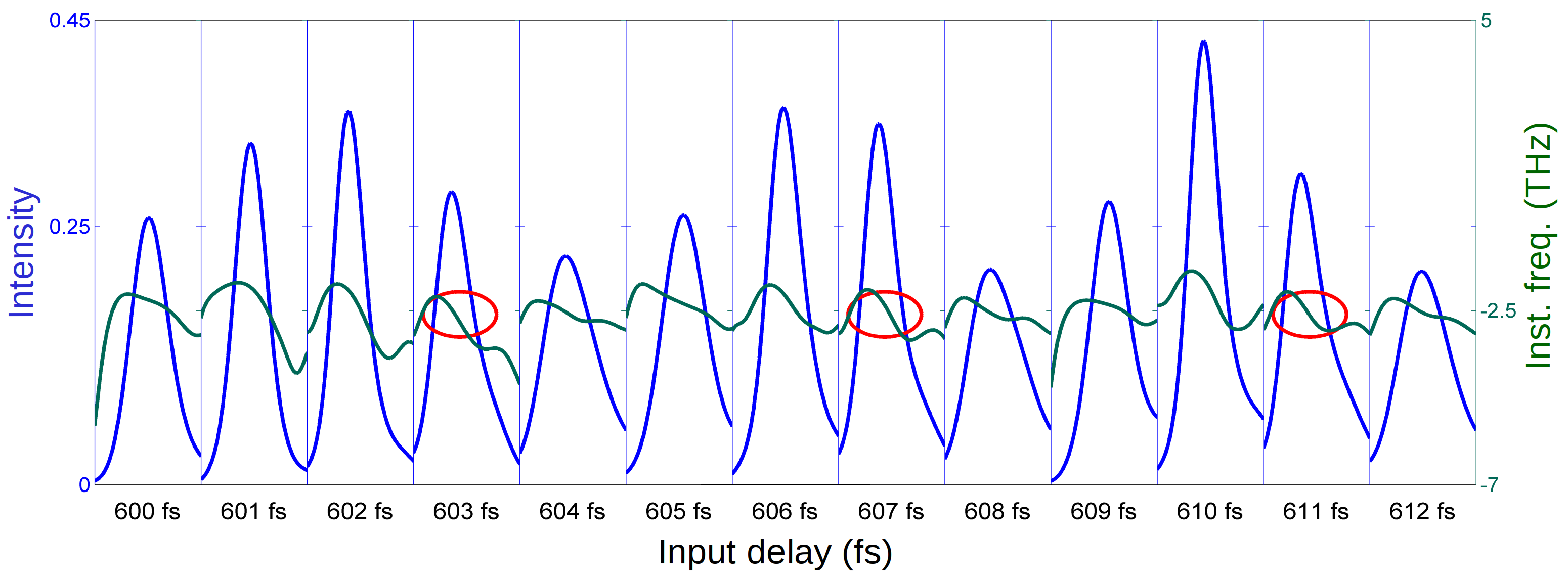}
	\caption{Measured Ramsey fringes. Measurements of intensity (blue traces) and instantaneous frequency (green traces) profiles for a delay range of $600-612~fs$. The intensity and instantaneous frequency are cyclical with a period of an optical cycle at $1.55~ \mu m$ ($ \sim 5 ~fs$). The instances when the instantaneous frequency repeats itself are marked by red circles.}
	\label{Ramsey_3}
\end{figure*}

The evolution of the probe pulse intensity and its instantaneous frequency profiles are shown in Fig. \ref{Ramsey_3}. The intensity oscillates with a period of $ \sim 5 ~fs$ and the modulation depth is $50 \%$. The instantaneous frequency profile is also cyclical with the same periodicity.  The instances of equal instantaneous frequency are marked by red circles.

\begin{figure}[htb]
	\centering
	\includegraphics[width=8.6cm]{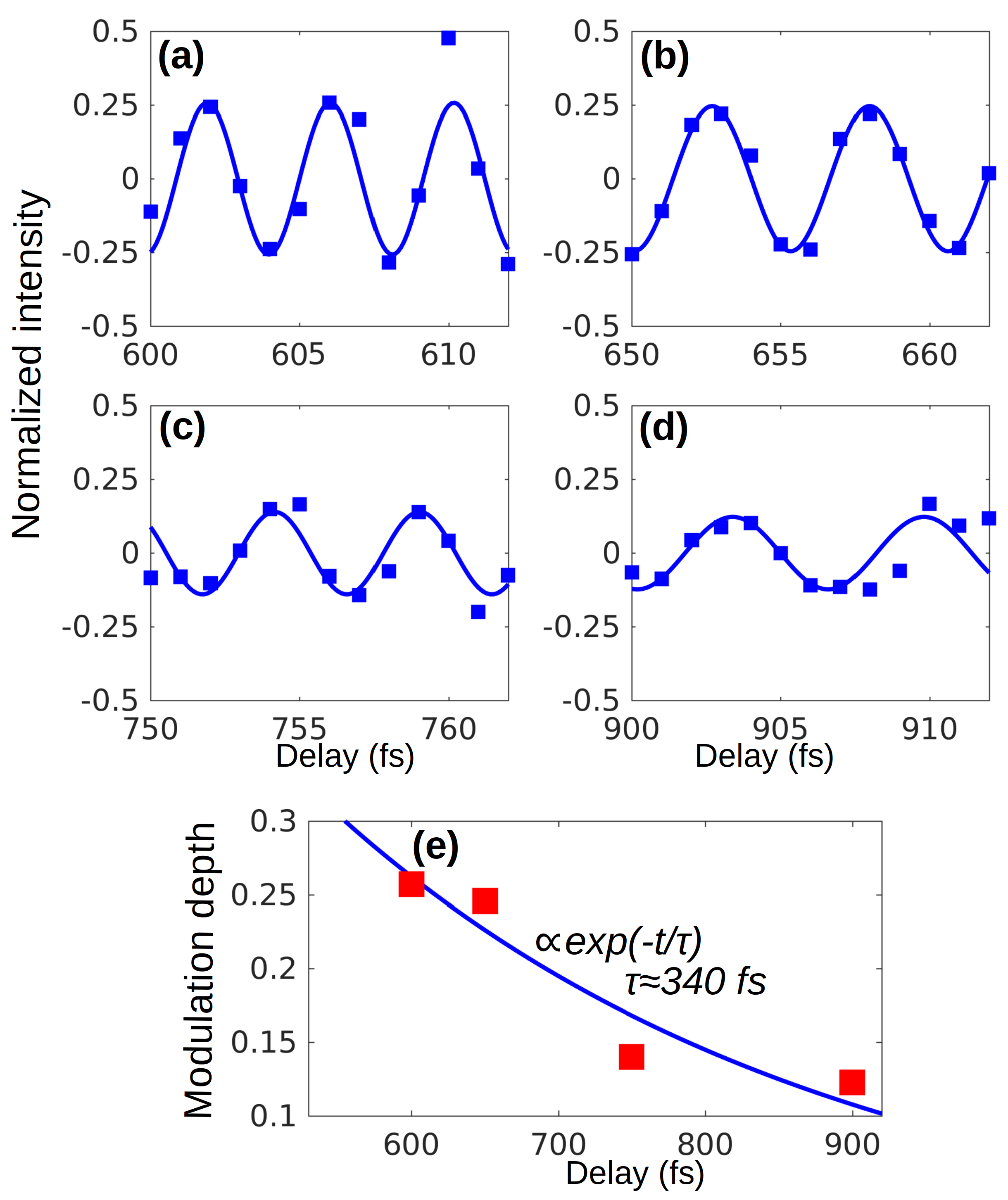} 		
	\caption{Normalized intensities of the probe pulse for various nominal delays. (a) $600~fs$, (b) $650~fs$, (c) $750~fs$, (d) $900~fs$. Each figure represents a $12~fs$ delay span. The decay of the modulation depth with nominal delay is clearly seen. It represents a direct mapping of the loss of coherence. (e) Modulation depth as a function of nominal delay from which a coherence time of $340~fs$ is extracted by a fit to an exponential function.}
	\label{Ramsey_4_A_D}
	\label{Ramsey_4_E}
\end{figure}

The dependence of the probe intensity oscillations on nominal input delay is described in Fig. \ref {Ramsey_4_A_D} (a)-(d), which shows the normalized intensity of the probe pulse for four sets of delays: $600$, $650$, $750$ and $900~fs$. An input delay of $600~fs$ is the shortest separation for which there is absolutely no input pulse overlap. This ensures that no interference occurs, which can mask the QD-mediated coherent coupling between the pulses. The intensity modulation depths are summarized in Fig. \ref{Ramsey_4_E} (e), which represents a direct mapping of the loss of coherence. A fit to an exponential decay (with an error of $R^2 \sim 0.9$) yields an extracted room temperature coherence time of $340~fs$.

\begin{figure}[htb]
	\centering
	\includegraphics[width=8cm]{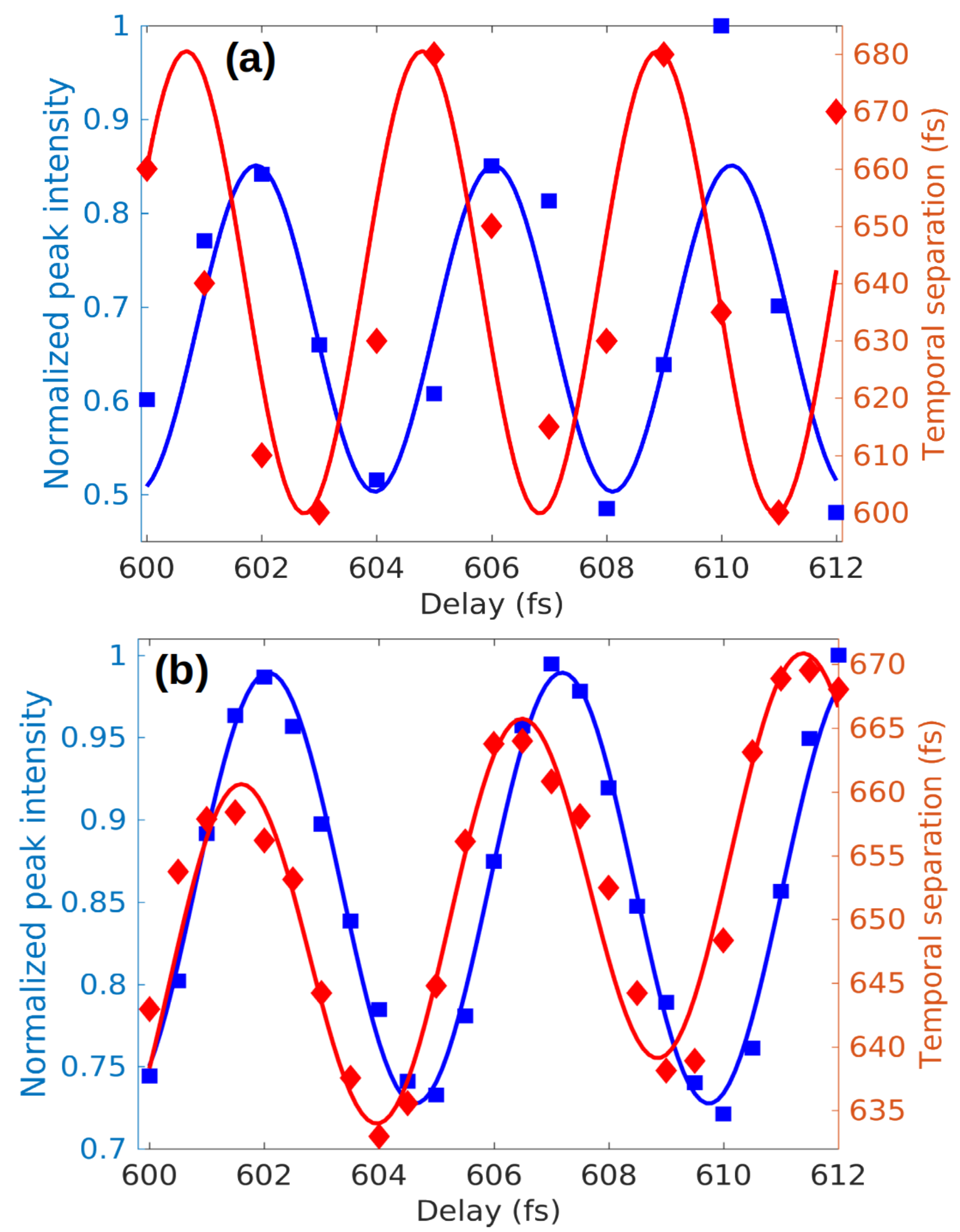} 	
	\caption{ Normalized intensity and output pulse separation for a delay range of $600$ to $612~fs$.	(a) Measured amplitude variations (blue trace) and output pulse separations (red trace). (b) Simulated results confirming the measurements shown in (a). The variation of the output pulse separation lags behind the intensity modulation by a quarter of a cycle due to the complex nature of the susceptibility.}
	\label{Ramsey_5_A}
	\label{Ramsey_5_B}
\end{figure}

The amplitude oscillations of the probe pulse are accompanied by a unique phenomenon - an oscillation in the output pulse temporal separation. The amplitude changes cause modifications of the carrier density which modulates, in turn, the real part of the susceptibility and hence the group velocity and the pulse propagation time. This is shown in Fig. \ref{Ramsey_5_A} (a) for delays of $600$ to $612~fs$ where the measured peak output intensity of the probe pulse is plotted as a blue trace and the output temporal separation between the peaks of the output pulses is shown in the red trace. Both exhibit a Ramsey-type fringe pattern with the same oscillation frequency but with a quarter cycle phase shift. The phase shift stems from the fact that changes in the real part of the susceptibility, induced by modulation of the imaginary part, translate in the time domain to a delay of a quarter of a cycle. The dependence on nominal delay of the output pulse separation follows a similar trend to that of the amplitude modulation depth namely, it decays exponentially as the delay increases.

A comprehensive finite difference time domain model was used to simulate the observed Ramsey fringes with the measured input pulses used as the excitation. The model treats the active medium of the amplifier as a cascade of effective two-level systems \cite{capua2013finite}. It calculates the co-evolution of the electromagnetic wave and the electronic state of the gain medium along the propagation axis of the amplifier by solving simultaneously Maxwell's and Schr$ \ddot{o}$dinger equations in the density matrix formalism. The model accounts also for the QD gain inhomogeneity as well as for non-resonant propagation effects \cite{karni2015nonlinear}. The QDs are fed incoherently from electrically driven common carrier reservoirs. Each confined electron state is associated with one excited level, which is energetically close to the carrier reservoir while the dense excited valance band states are treated as one effective state, merged into the hole reservoir. The simulation results are shown in Fig. \ref{Ramsey_5_B} (b). The oscillations of the peak intensity as well as the pulse separation confirm the measured results.

To conclude, we have used a high resolution two-pulse pump-probe-X-FROG system to demonstrate Ramsey interference in an electrically driven $1.55~ \mu m$ InAs/InP QD SOA operating at room temperature. The probe intensity and its instantaneous frequency profiles exhibit a cyclical behavior with a periodicity of $ \sim 5 ~fs$, which equals a single optical cycle at $1.55~ \mu m$. The modulation depth in the oscillating probe intensity decreases with increasing nominal input delay and this enables a direct mapping of the de-coherence process that the active gain medium undergoes and the extraction of a coherence time which is found to be $340~fs$. Above and beyond this, we also demonstrated an unique property by which the output separation between the two pulses also oscillates with the same periodicity and exhibits a delay of one quarter of a cycle relative to the intensity oscillation. The quarter cycle delay is due to the coupling between the two parts of the complex susceptibility. The oscillation in pulse separation is only possible in a distributed medium such as the QD SOA and cannot occur in the single QD systems used in many Ramsey interference experiments. The experimental results were confirmed by a comprehensive numerical model.
\section*{Funding Information}
Israel Science Foundation (1504/16).

\section*{Acknowledgments}
IK acknowledges Dr. Elena Khanonkin, Mr. Ori Eyal, Mr. Shai Tsesses for helping in performing the measurements. AKM acknowledges the support in part at Technion by Israel Council for Higher Education.

\bibliography{sample}

\end{document}